\documentclass[10pt]{iopart}

\usepackage{amsfonts}
\usepackage{amssymb}
\usepackage{xfrac}
\usepackage{graphicx}
\usepackage{booktabs,array}
\usepackage{comment}
\usepackage[skip=2pt,font=footnotesize]{caption}
\usepackage{titlesec}
\usepackage[flushleft]{threeparttable}
\usepackage{color}
\usepackage[normalem]{ulem}
\usepackage{tablefootnote}
\usepackage[backend=bibtex,sorting=none,style=numeric,natbib=true]{biblatex}

\newcolumntype{L}[1]{>{\raggedright\arraybackslash}m{#1}}
\newcolumntype{C}[1]{>{\centering\arraybackslash}m{#1}}
\newcolumntype{R}[1]{>{\raggedleft\arraybackslash}m{#1}}

\titleformat*{\section}{\normalsize\bfseries\rmfamily}
\titleformat*{\subsection}{\normalsize\bfseries\rmfamily}
\titleformat*{\subsubsection}{\normalsize\bfseries\rmfamily}

\graphicspath{{./Figures/},{./Workflows/}}
\DeclareGraphicsExtensions{.eps,.pdf,.jpg,.png}
\bibliography{library}

\begin{document}

\title{First-principles-based multiple-isotope particle transport modelling at JET}

\author[M. Marin, J. Citrin, C. Bourdelle, Y. Camenen, F. J. Casson, A. Ho, F. Koechl, M. Maslov and JET Contributors]{M. Marin$^1$, J. Citrin$^1$, C. Bourdelle$^2$, Y. Camenen$^3$, F. J. Casson$^4$, A. Ho$^1$, F. Koechl$^4$, M. Maslov$^4$ and JET Contributors$^*$}

\address{$^1$ DIFFER - Dutch Institute for Fundamental Energy Research, Eindhoven, the Netherlands}
\address{$^2$ CEA, IRFM, F-13128 Saint-Paul-lez-Durance, France}
\address{$^3$ CNRS, Aix-Marseille Univ., PIIM UMRT7345, 13397 Marseille Cedex 20, France}
\address{$^4$ CCFE, Culham Science Centre, Abingdon, OX14 3DB, United Kingdom of Great Britain and Northern Ireland}
%\address{$^5$ OAW/ATI, Atominstitut, TU Wien, 1020 Vienna, Austria}
\address{$^*$ See Joffrin et al 2019 (https://doi.org/10.1088/1741-4326/ab2276) for the JET Contributors.}

\begin{abstract} 

Core turbulent particle transport with multiple isotopes can display observable differences in behaviour between the electron and ion particle channels. Experimental observations at JET with mixed H-D plasmas and varying NBI and gas-puff sources [M. Maslov \textit{et al.,} \textit{Nucl. Fusion} \textbf{7} 076022 (2018)] inferred source dominated electron peaking, but transport dominated isotope peaking. In this work, we apply the QuaLiKiz quasilinear gyrokinetic transport model within JINTRAC flux-driven integrated modelling, for core transport validation in this multiple-isotope regime. The experiments are successfully reproduced, predicting self consistently $ j $, $ n_{e} $, $ n_{Be} $, $ T_{e} $, $ T_{i} $, $\omega_{tor}$ and the isotope composition. As seen in the experiments, both H and D profiles are predicted to be peaked regardless of the core isotope source. An extensive sensitivity study confirmed that this result does not depend on the specific choices made for the boundary conditions and physics settings. While kinetic profiles and electron density peaking did vary depending on the simulation parameters, the isotope ratio remained nearly invariant, and tied to the electron density profile. These findings have positive ramifications for multiple-isotope fuelling, burn control, and helium ash removal.

\end{abstract}

\maketitle
\ioptwocol

\section{Introduction} \label{Introduction}

Understanding and predicting particle transport is essential to interpret and optimize experiments \cite{Angioni2009}. Due to ambipolarity, the total ion and electron transport must be equal, and it is experimentally challenging to separate ion and electron transport coefficients. Differences between the ion and electron particle channels can only be observed in a multiple ion plasma. These differences will be studied in this paper. Since the plasma density diagnostics primarily focus on the electron density profile, most of the attention on this topic has historically been on the electron particle transport. The particle flux $ \Gamma_{s} $ is usually formalized as 

\begin{equation}
    \Gamma_{s} = D_{s} \frac{dn_{s}}{dr} + n_{s}V_{s}
\end{equation}

with $ D_{s} $ the diffusion coefficient for the species $ s $ and $ V_{s} $ the convective term. Both have a neoclassical \cite{Hinton1976} and a turbulent component \cite{Zabolotsky2003}, the latter usually larger than the former \cite{Gentle1987, Wagner1993}.

Turbulent convection was proven experimentally \cite{Garbet2003, Hoang2006} and theoretically \cite{, FJenkot2005, CAngioni2009c, CEstrada-Mila2006} in absence of core particle source and of neoclassical ware pinch \cite{Ware1970}. This followed from observations of higher density in the core than at the edge (L-modes) and than at the top of the pedestal (H-modes) and is commonly referred as density peaking. In the absence of a core particle source it is caused by an inward convection, also called particle pinch, and depends on the turbulent regime \cite{Fable2010}. When a particle source is present, the contribution to the density peaking from source and convection varies depending on the conditions \cite{Tala2019}.

The basic mechanisms for density peaking can be interpreted through the continuity equation, written in cylindrical coordinates for simplicity of exposition:

\begin{equation}
\label{eq:continuity}
    \begin{array}{l}
        \frac{\partial n_{s}}{\partial t} - \frac{1}{r}\frac{\partial}{\partial r}(r\left(D_{s}\frac{\partial n}{\partial r}-V_{r}n\right)) =  S_{s}
        \end{array}
\end{equation}

Where $s$ denotes a plasma species, whether electrons or ions. Supposing stationary state, multiplying by $r$ and carrying out volume integration, we obtain

\begin{equation} \label{eq:simplified_continuity}
    -\int_{0}^{r}\frac{\partial}{\partial r}(rD_{s}\frac{\partial n}{\partial r})dr =   -\int_{0}^{r}\frac{\partial}{\partial r}rV_{s}ndr + \int_{0}^{r}S_{s}rdr
\end{equation} 

And therefore, after some algebra

\begin{equation} \label{eq:density_peaking}
    -\frac{1}{n_{s}}\frac{\partial n_{s}}{\partial r} = -\frac{V_{s}}{D_{s}} + \frac{1}{n_{s}rD_{s}}\int_{0}^{r} r' S_{s} dr'
\end{equation}

If the $-V_s/D_s$ term on the RHS dominates, then the density peaking for that species is controlled by transport. The importance of the source term (second term on RHS) depends on the magnitude of the particle diffusivity. A large $ D_{s} $ implies a weak impact of the source on the density peaking. This would at the same time, for a peaked density, imply a large pinch (inward convective term $V_s$) \cite{Angioni2009a, Cardoso1995}.

While in a single-ion plasma there cannot be a difference between electron and ion transport, this is not as trivial when multiple ions are considered, since they can mix without affecting the electron density. Experimentally, fast isotope mixing with trace-T was observed both in TFTR \cite{Efthimion1995} and at JET \cite{Zastrow2004}. Large He transport with $D_{He}$ on the order of $\chi_i$ was observed in AUG by both modelling and experiments \cite{Angioni2009a}.

Recent experiments at JET were performed with two main ion species, Hydrogen (H) and Deuterium (D) \cite{Maslov2018}, with the explicit goal of understanding the impact of the core particle source on the different isotope profiles. The isotope sources were varied by scanning the relative contribution of peripheral gas injection (edge source) and neutral beam injection (core source). The edge composition was measured comparing the relative amplitude of Balmer $ H_{\alpha} $ and $D_{\alpha} $ spectral lines, while the Deuterium density in the deep core was derived from the neutron rate.

It was found in these experiments that for both isotopes the density was peaked. This was observed regardless of whether the core isotope source was purely Deuterium or purely Hydrogen. This work focuses on investigating and interpreting these observations with first-principle-based quasilinear turbulent transport modelling within a flux-driven integrated modelling framework.

For the pulses studied in this work, analysis outlined in Ref.~\cite{Maslov2018} have indicated a disparity in magnitude between the ion and electron particle transport coefficients, with $D_{i}/D_{e} \gg 1$. 

Regarding $D_e$, from studies on the JET density peaking database \cite{Angioni2007b}, transport models \cite{Garzotti2003a, Garzotti2006}, and recent gas puff modulation experiments on DIII-D \cite{Mordijck2015}, a strong correlation was found between the effective electron collisionality $ \nu_{eff} $ and the density peaking. The pulses studied in this work have $ \nu_{eff} \sim 0.1 $. The degree of density peaking associated with this $\nu_{eff}$ value from previous studies is consistent with the observed values here. The electron particle diffusion coefficient $D_e$ is then expected to lie within $ 0.2 \chi_{eff} < D_{e} < 0.6 \chi_{eff} $. A $ D_{e} $ in this range is expected to lead to a significant contribution from the 8MW NBI to the density peaking. While other parameters beyond $\nu_{eff}$ can also have an impact on the density peaking and $D_e$, this observation hints towards small electron transport coefficients.

Regarding $D_i$, a stationary state methodology in Ref. \cite{Maslov2018} (no gas puff modulation), only applicable for multi-ion plasmas, was applied to estimate the $D_i$, $V_i$ of the D and H isotopes. At mid-radius, $ D \approx 3.6 [m^{2}s^{-1}] $ and $ V \approx -2.2 [ms^{-1}]$. At the same position, $ \chi_{eff} \sim 2[m^{2}s^{-1}] $. This implied $ D_{i}/\chi_{eff} \sim 1.5 $, underlying the disparity with the expected $D_e$.

Following this experimental work, theory and modelling showed that $D_i/D_e \gg 1$ and $|V_i|/|V_e|\gg1$ holds for Ion-Temperature-Gradient (ITG) dominated turbulence \cite{Bourdelle2018a}. Ambipolarity is maintained by the large ion diffusion balanced by a large inward pinch. This was observed consistently in Ref.\cite{Bourdelle2018a} by a quasilinear analytical model showing the provenance of particle transport coefficient magnitudes in wave-particle resonances, with nonlinear GKW \cite{Peeters2009} gyrokinetic simulations, and by the quasilinear gyrokinetic turbulent transport model QuaLiKiz \cite{bourdelle2016,Citrin2017}. Validation of this effect in QuaLiKiz compared to nonlinear simulations is pertinent for this work, since QuaLiKiz is applied here for flux-driven validation against the experiments.

The resonant nature of this effect changes depending on the dominant underlying modes driving the turbulence. The effects are opposite for pure ITG and Trapped-Electron-Mode (TEM) turbulence. In the TEM regime, the electron transport particle coefficients are in fact larger than the ion ones. The aforementioned experiments at JET were found to be purely ITG dominated on ion scales, so in line with the theoretical predictions.

Qualitative 'numerical experiments' had already been performed in \cite{Bourdelle2018a}, showing peaked isotope density profiles regardless of which isotope was used in the NBI. This work focuses instead on quantitatively reproducing the main experimental results in Ref. \cite{Maslov2018} within the framework of integrated modelling, using the JINTRAC suite \cite{Romanelli2014} with QuaLiKiz as the turbulent transport model. We show that we can reproduce the experimental temperature and density profiles, which provide confidence that the correct ITG regime is being captured. We then show that we obtain quantitative agreement with the experimental Deuterium profiles and that the phenomena is robust and quite insensitive to the boundary conditions and physical hypotheses in the model.

The studied pulses are described in \sref{pulses}. The tools and methods are discussed in \sref{Tools_methods}. In \sref{Assumptions} the simulations setup is reviewed. The results of the analysis will be summarized in \sref{Results} and an overview of the sensitivities of the model will be presented in \sref{Sensitivities}. Conclusions are drawn in \sref{Conclusions}

\section{Experimental discharges} 
\label{pulses}

The focus of this work is on three discharges performed with mixed H and D isotopes during the JET experimental campaign in 2016, two of which are extensively described in Ref.\cite{Maslov2018}. All the pulses have 8 MW of injected Neutral Beam Injection (NBI) heating power, plasma current $ I_{p} = 1.38 MA $, and magnetic field $ B_{t} = 1.7 T $. The details of the discharges and the core D concentration are reported in \tref{tab:pulses_details}.

\begin{table*}
\caption{Details of the studied pulses. Core source is the NBI, edge source is the gas puff. The edge composition is inferred from the intensity of Balmer $ \alpha $ lines, the core composition from the neutron rate. $ \langle T_{e} \rangle $ and $ \langle n_{e} \rangle $ are the volume averaged electron temperature and density. There was no precise measure of the core composition for 91754. The * emphasizes that the core composition for 91232 is the one chosen in Ref.\cite{Maslov2018} using a conservative approach which corresponds to $ +\sim 10 \% $ error in the neutron rate.}
\begin{indented}
	\centering
	\item[]\begin{tabular}{cccccccccc}
		    \br 
		    Pulse $ \# $ & \vtop{\hbox{\strut Averaged}\hbox{\strut between [s]}} & \vtop{\hbox{\strut Core}\hbox{\strut Source}} & \vtop{\hbox{\strut Edge}\hbox{\strut Source}} & $Z_{eff} $ & Shift [cm] & $ \frac{n_{D}}{n_{H} + n_{ D}} $ $ \rho = 0.8 $ & $ \frac{n_{D}}{n_{H} + n_{D}} $ $ \rho = 0 $ & $ \langle n \rangle [10^{19}m^{-3}] $ & $ \langle T_{e} \rangle [KeV] $ \\ 
    		\mr
    		91754 & 6.4 - 7 & H & H+D & 1.15 & 2.9 & 0.53 & - & 2.6 & 1.05 \\ 
    		
    		91232 & 5 - 6 & D & H & 1.2 & 3.1 & 0.15 & 0.18* & 2.4 & 1.05 \\ 
    		
    		91227 & 8.2 - 8.5 & D & H+D & 1.2 & 3.3 & 0.64 & 0.676 & 2.7 & 1.05 \\ 
    		\br
    	\end{tabular}
    	\label{tab:pulses_details}
	\end{indented}
\end{table*}

\begin{figure*}
	\centering
	\includegraphics[width=1.0\linewidth]{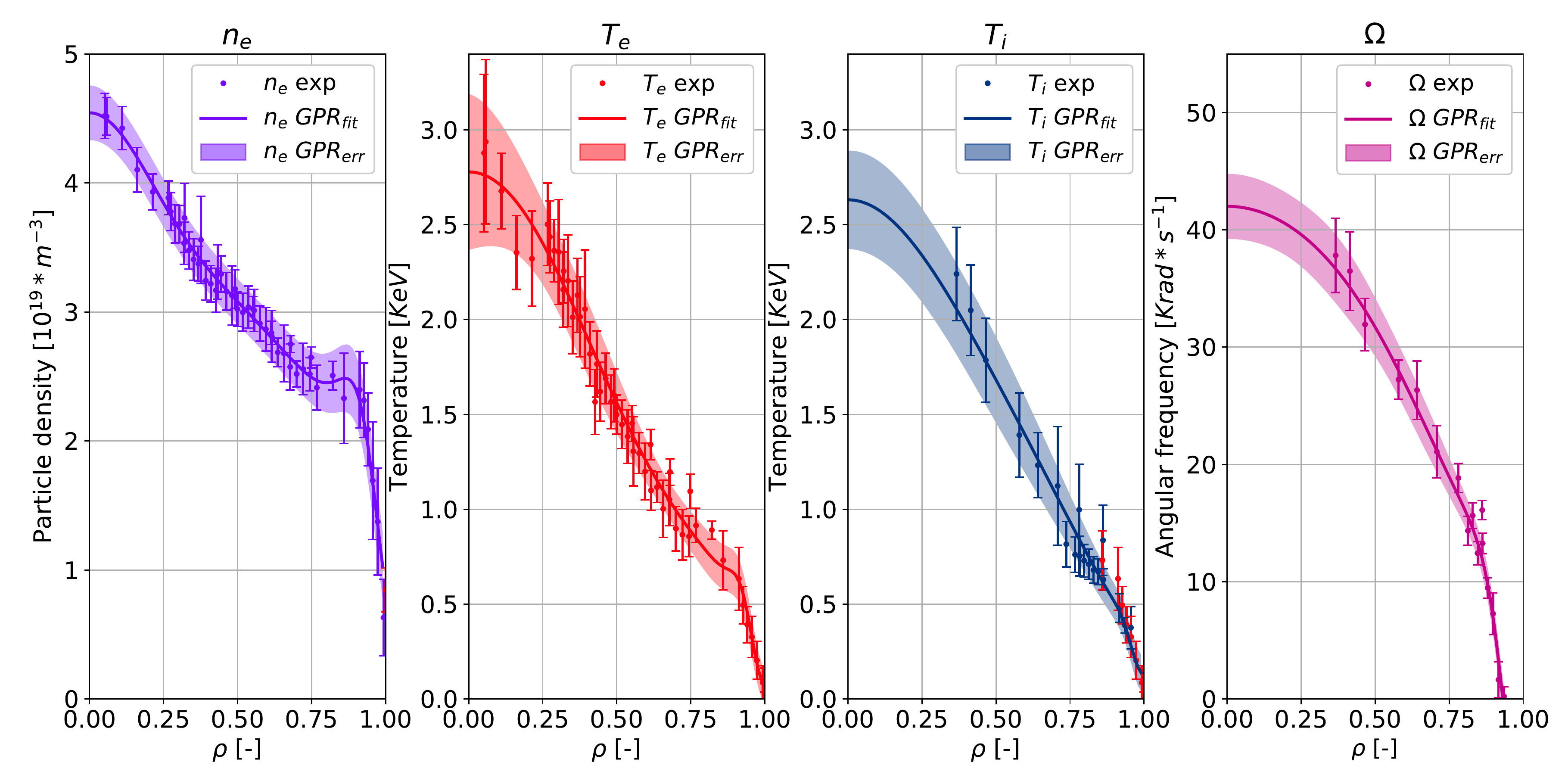}
	\caption{Experimental data and Gaussian Process Regression (GPR) fits for 91754. The dots are the experimental data, HRTS for density and electron temperature, edge and core CX for ion temperature and rotation. The solid lines are the GPR fits, with the $ 1\sigma $ confidence interval showed by the shaded area. The red points in the third plot are HRTS $ T_{e} $ data added due to the the scarcity of $ T_{i} $ data in the pedestal region, where $ T_{i} \sim T_{e} $ is assumed.}
	\label{fig:91754_fits}
\end{figure*}

The experimental profiles were fitted using a Gaussian Process Regression (GPR) tool available at JET \cite{ho2019}, employing a Gibbs kernel. GPR has the advantage of providing both fit and derivative uncertainties, making it suited for model verification and validation. The fitted profiles are employed as the initial condition for the integrated modelling simulations. The electron temperature and density were obtained from High Resolution Thomson Scattering (HRTS). Ion temperature and rotation were obtained from edge and core charge exchange (CX). To account for equilibrium reconstruction uncertainty, the kinetic profiles were radially shifted to set $ T_{e} \sim 100eV $ at the last-closed-flux-surface (LCFS), as standard in pedestal modelling at JET \cite{fras2019}. A further constraint was added increasing the shift if the density at the separatrix resulted to be lower than $ 0.4 \langle n_{e} \rangle $ \cite{Kallenbach2005} . The magnitude of the shift is reported in \tref{tab:pulses_details}.

Due to technical issues with one of the HRTS spectrometers, HRTS data between $ 0.85 < \rho < 0.9 $ was not available for the discharges studied. $ \rho $ here is the normalised toroidal flux coordinate $ \rho_{tor} = (\frac{\psi_{tor}}{\psi_{tor, LCFS}})^{\frac{1}{2}} $.
Edge CX was only available for $\#91754$, providing a value for $T_{i}$ very similar to $T_{e}$ close to the pedestal top (\fref{fig:91754_fits}).

The isotope composition at the edge was measured by comparing the relative amplitude of Balmer $ H_{\alpha} $ and $ D_{\alpha} $ spectral lines. Good agreement was obtained between the two diagnostics that were were used, one looking at the plasma edge and the other one using a penning gauge discharge to measure the composition of the sub-divertor neutral gas. The core composition was calculated from the neutron rate. For $\#91232$ and $\#91227$ most of the neutrons come from D - D Beam - Thermal reactions. Less than $ 0.5\% $ of the neutrons come from the Thermal - Thermal channel, which is ignored.

From TRANSP analysis, around  80 $\%$ of the neutrons originate from the $ \rho < 0.5 $ region. After subtracting the non negligible Beam - Beam contribution, around 25$\%$ in 91232, the neutron rate is used to constrain the deep-core Deuterium density through TRANSP/NUBEAM modelling \cite{Maslov2018}.

The experimental data were averaged for $ \sim 400 ms $ during flat - top, as defined by kinetic profile quasi-stationarity over several confinement times. The intervals include several ELMs and a few sawteeth. The GPR simply averages the experimental data and this typically results in larger error bars close to the axis and to the pedestal region. A stationary state was also reached in the integrated modelling simulation by letting the profiles relax for 2s, corresponding to $ \sim 10 $ (electron) particle confinement times.

\section{Tools and methods} \label{Tools_methods}

Multi-channel integrated modelling is a powerful tool for transport model validation, experimental analysis and interpretation, and "Predict First" applications for scenario prediction, design and optimization. The key aspect is constraining the kinetic profiles to power and particle balance, and capture the complex nonlinear interactions between multiple transport channels, sources, sinks, and magnetic equilibrium. JINTRAC was the framework chosen for this work, with JETTO \cite{Jetto_manual} as the transport solver. Regarding the particle transport, which is central in this paper, it is important to specify that JINTRAC evolves only the ion densities. The electron density is then set by quasineutrality.

One simulation henceforth named 'Basecase' was performed for each shot with the settings that were presumed to be correct. These settings will be presented below. For every choice the impact was assessed, allowing the simple estimates to be justified and identifying the important modelling knobs.

The modelling was constrained to the core, with a boundary condition taken at $ \rho = 0.8 $ from the GPR fit, which is unambiguously inside the pedestal region. Between $ 0.2 < \rho < 0.8 $, we self-consistently predicted the current ($ j $), electron and ion temperature ($T_{e}, T_{i}$), the electron, Hydrogen, Deuterium and Beryllium density ($n_{e}, n_{H}, n_{D}, n_{Be}$) and the rotation ($\Omega $). The neoclassical and turbulent transport were calculated with NCLASS \cite{Houlberg1997} and QuaLiKiz respectively. The levels of radiation were always low ($P_{rad} \simeq 1MW $), implying a low level of tungsten (W) concentration. Therefore it was decided not to model W, as was done for example in \cite{Breton2017, CassonIAEA}, since it would have required the use of a higher fidelity and time-consuming neoclassical transport model. 

The neutral source at the edge was calculated with FRANTIC \cite{Tamor}, the NBI heat and particle sources with PENCIL \cite{Challis1989}, the impurity transport evolution with SANCO \cite{Alper}, the magnetic equilibrium with ESCO \cite{Jetto_manual}. Sawteeth are not modeled in a fully predictive manner, since this slows down and unduly complicates the simulations. The inversion radius is in all cases around $ \rho \sim 0.3 $. Within this radius, we simply added additional transport as a proxy for sawteeth. The additional transport is added as a Gaussian with amplitude respectively 0.25, 0.5, 1.2 and 0.7 $ m^{2}s^{-1} $ for particle, electron thermal, ion thermal and momentum diffusivities.

It is worth noting that the simulations for all 3 discharges studied here apply the same settings, with no case-dependent fitting parameters.

\section{Modelling assumptions} \label{Assumptions}

There are multiple aspects in the setting up of the simulations that require special care. The decisions made in this process are presented below.

\subsection{Boundary conditions} \label{Boundary}

As will be explored more later, the values of density and temperature at the internal boundary condition ($ \rho = 0.8 $) have a significant influence on the simulated profiles. This means that special attention is necessary for the experimental values in this region.

The $T_i/T_e$ ratio at the internal boundary condition was found to have a considerable impact on the density peaking in QuaLiKiz simulations \cite{Linder2019,ho2019}. An incorrect estimation of $T_{i}$ at our $ \rho = 0.8 $ boundary condition could thus have an indirect deleterious effect on the density peaking prediction.
In such relatively low power - high collisionality discharges, $T_{i}$ is usually assumed to be very similar to $T_{e}$. Since $T_{i} = T_{e} $ is within the error-bars in these discharges, especially in $ \#91754 $ where edge CX is available, $T_{i} = T_{e}$ at $ \rho = 0.8 $ was imposed for all the discharges.

\subsection{Sources} \label{Sources}

PENCIL was used for the NBI electron and ion heat sources, the torque source and the core particle source. The total injected power was 8MW. The beam energy fractions were set to be consistent with the experiment, while the ion energy was averaged between the ion energies of the active PINIs.

The edge neutral source was calculated by FRANTIC. The ionization energy per atom and the wall released neutral energy were set to 14eV and 300eV respectively. A penetration efficiency of the gas puff into the LCFS of $ \sim 10\% $ percent was used, similar to what was found for tritium puff in Ref. \cite{Zastrow2004}. A standard method to measure the efficiency is not available, but some assumptions can be made. In similar conditions, it was found that the gas puff edge particle source was similar to the NBI particle source around mid radius \cite{Valovic2004}. In our simulations, the edge particle source is equal to the NBI particle source around $ \rho = 0.7 $. Doubling the puff in FRANTIC moved the position where the two sources are equal at $ \rho = 0.6 $ and had a negligible impact on the core Deuterium concentration.

\subsection{Equilibrium} \label{Equilibrium}

The kinetic profiles are dependent on the q-profile primarily through the $s/q $ ratio, which has a strong impact on the ITG and ETG thresholds \cite{Guo1993,Jenko2001, Citrin2012}. An accurate equilibrium reconstruction is therefore mandatory. It was decided to calculate the magnetic equilibrium self-consistently via current diffusion simulations and ESCO solutions to the Grad Shafranov equation. The experimental internal inductance change was observed to be comparable with the noise level ($\sim5 \% $) for a timescale larger than the current evolution timescale, so the current profile was assumed to be relaxed.
The initial q-profile was obtained from EFIT \cite{Search1985} and then evolved in a simulation with fixed temperature and density profiles but predictive current. Since in the experiment sawteeth were present, the sawteeth model in JETTO was used in these preparatory simulations. The crash times were calculated by the Porcelli model \cite{Porcelli1996}, while a simple Kadomstev model was used as the re-connection model \cite{Kadomstev1975}. The relaxed profiles had an inversion radius around $ \rho \sim 0.3 $, similar to the experimental value.
This relaxed q-profile was the one chosen for each Basecase. The internal inductance in the experiments is around $ li \approx 0.88 $ in all cases, while in the model we obtain $ li \approx 0.92 $. This slight discrepancy could be due to differences in the equilibrium, in the kinetic profile fits or in the $ Z_{eff} $ profile, impacting both the resistivity profile and the bootstrap current. 

%The precise parameters in the sawteeth model were found to have a negligible impact on the final current profile, so the parameters presented in \cite{Porcelli1996} for ITER standard operation were used.

\subsection{Impurities} \label{Impurities}

The Beryllium impurity was modelled predictively with SANCO, with the transport coefficients calculated by NCLASS and QuaLiKiz. ADAS96 \cite{Summers1994} was used for the atomic data. Being $ Z_{eff} \sim 1.2$ for all discharges, only Beryllium (Be) was used, since it is the largest contribution to the main ion dilution. Tungsten (W) contribution to dilution and $ Z_{eff} $ is almost negligible in these discharges, but is the main source of radiation. Since radiation from the core is less than 1MW in all cases, the impact on the profile evolution is negligible. W-modelling requires higher fidelity neoclassical models which slow down the simulations, so it was decided to not include W. A flat radiation profile in the core was assumed.

\subsection{QuaLiKiz settings} \label{QuaLiKiz}

Electron Temperature Gradient (ETG) driven turbulence was found to carry a modest part of the electron heat flux in all cases, so the electron scales were always included with the saturation rules proposed in \cite{Citrin2017b}.

The impact of rotation on the turbulence was included for $ \rho > 0.5 $, due to the under-estimation of destabilizing parallel-velocity-gradient drive in QuaLiKiz, more important in the inner half radius \cite{Citrin2017b}

Fast ion species were not included in the integrated modelling, to save computation time. From standalone (electrostatic) QuaLiKiz analysis, their impact was seen to be negligible.

\section{Simulations Results} \label{Results}

The Basecases, i.e. the simulations with the default settings outlined in section \ref{Tools_methods}, were compared with the experimental data and with the GPR fits. 

\begin{figure*}
	\centering
	\includegraphics[width=1.0\linewidth]{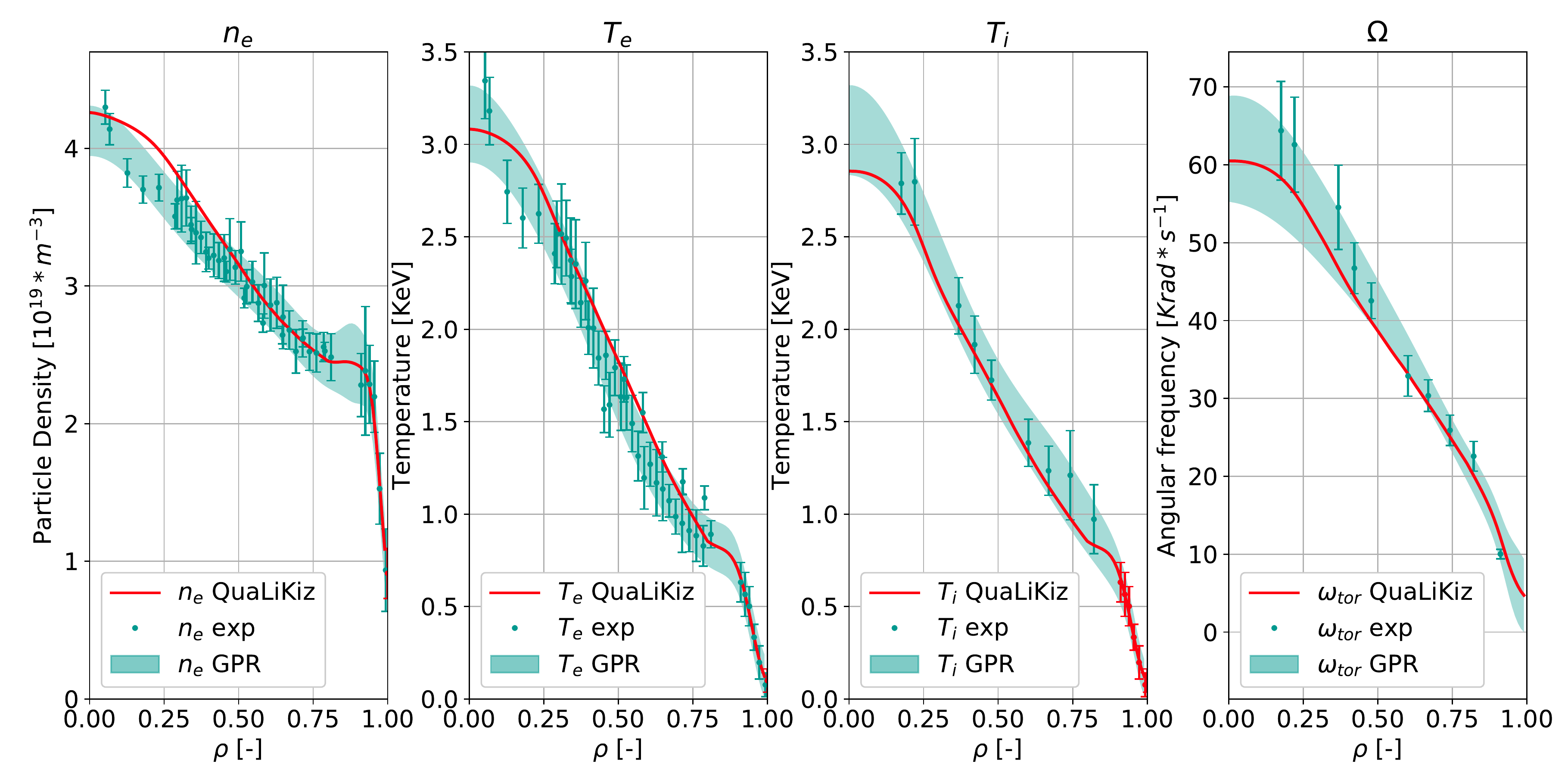}
	\caption{Comparison between the modelled profiles and the experimental data for $ \#91227 $ (D beam, mixed puff). The green points are the experimental data, HRTS for the density and the electron temperature, core CX for the ion temperature and rotation. The shaded areas are the $ 1\sigma $ confidence interval given by the GPR fits. The red lines are the profiles calculated by the integrated modelling, showed at the end of the simulations, when all the profiles are relaxed.}
	\label{fig:91227}
\end{figure*}

The results of the simulations for 91227 are shown in \fref{fig:91227}. The full comparison over all the channels is not shown for all the pulses, but the agreement is summarized in \tref{tab:experiment_vs_model}. The figure of merit described in \cite{Citrin2017b} is used. To emphasize that the experimental data are not equidistant in $ \rho $ a summation is used instead of the integral for the comparison with the raw data. An integral is kept when the same figure of merit is used to compare the simulation with the GPR fits.

\begin{table*}
\caption{Standard deviation figures of merit for all the pulses. $ f_{sim} $ is the simulated quantity, $ f_{exp} $ is the measured quantity and $ f_{fit} $ is quantity obtained from the GPR fits. $ \rho_{min} $ is the minimum value of $ \rho $ considered. $ \rho_{min} = 0.2 $ is chosen here since for $ \rho_{min} < 0.2 $ the transport is dominated by the artificially added extra transport.}
    \begin{indented}
    	\centering
    	\item[]\begin{tabular}{ccccc}
    		\br 
            Pulse $ \# $ &  \multicolumn{4}{|c}{$ (\int_{\rho_{min}}^{\rho_{BC}}dx(f_{sim} - f_{fit})^{2})^{\frac{1}{2}}/(\int_{\rho_{min}}^{\rho_{BC}}dxf^{2}_{fit})^{\frac{1}{2}} $}\\
    	    \mr
        	 & $ \sigma_{n_{e}} $ & $ \sigma_{T_{e}} $ & $ \sigma_{T_{i}} $ & $ \sigma_{\Omega} $ \\ 
    		\mr
    		91754 (H beam, mixed puff) & 8.4\% & 6.6\% & 4.6\% & 15.9\% \\ 
    		
    		91232 (D beam, H puff) & 7.6\% & 3.4\% & 17.2\% & 4.8\% \\ 
    		
    		91227 (D beam, mixed puff) & 4.9\% & 7.1\% & 5.3\% & 5.2\% \\ 
    		\br
    	\end{tabular}
    	\label{tab:experiment_vs_model} 
	\end{indented}
\end{table*}

\begin{figure*}
	\centering
	\includegraphics[width=1.0\linewidth]{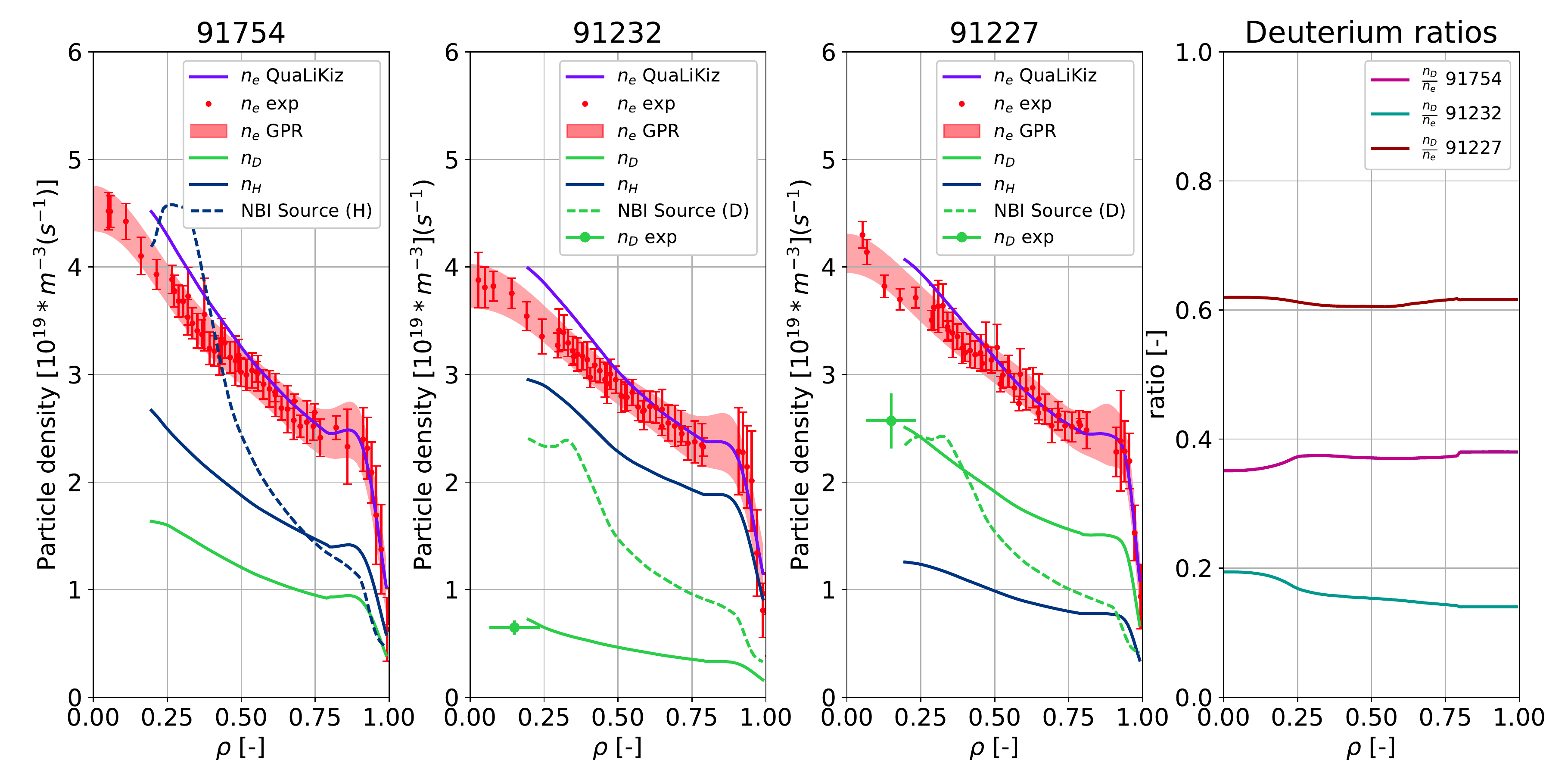}
	\caption{Electron and isotopes density profiles for $ \#91754, \#91232, \#91227 $. The red points are the experimental data from HRTS, the red shaded area is the $ 1\sigma $ confidence interval of the GPR fit, and the purple line is the model. Deuterium and Hydrogen modelled profiles are shown in green and dark blue. The dotted line is the core particle source, Hydrogen in blue and Deuterium in green. The $ n_{D}/n_{e} $ ratio is plotted for all the pulses in the rightmost plot.}
	\label{fig:Density_profiles}
\end{figure*}

The electron density profile modelled for the various cases are shown in figure \ref{fig:Density_profiles}, together with the modelled Hydrogen and Deuterium profiles. Crucially, the density is peaked for both isotopes regardless of the core source, as anticipated and observed in the experiment. The Deuterium to electron ratios for all 3 discharges are summarized in the rightmost plot. While the total Deuterium concentration can depend on the total fuelling and edge physics, the isotope composition ratio remains roughly constant in the core regardless of core isotope source.

This central result comes out naturally from the simulations. Both isotope density profiles follow the peaking of the electron density profile. To show that the interpretation of large ion diffusion and pinch coefficients is correct we plot the ratio between the QuaLiKiz predicted electron and ion particle transport coefficients for the last timestep of $ 91227 $.

Since in 91754 a Hydrogen beam was used, all the neutrons came from the thermal - thermal channel. The neutron rate is therefore very sensitive to the ion temperature and an accurate calculation of the isotope composition is difficult. 
Nevertheless, preliminary experimental analysis tried to set a maximum for the ion temperature and calculated the Deuterium core concentration, indicating a peaked Deuterium density profile. The model consistently predicted a peaked Deuterium profile, in spite of Hydrogen NBI and thus solely a Hydrogen core particle source.

While for 91227 very good agreement is reached, for 91232 $ \frac{n_{D}}{n_{H} + n_{D}} $ is slightly overestimated by JETTO - QuaLiKiz. We note that the overestimation comes from inside the Sawteeth inversion radius, where additional transport is artificially added and the model is no longer first-principles-based ( Table \ref{tab:experiment_vs_model_density}). The experimental error is estimated around $ 10\% $. Between the experimental and the modelling errors we can consider the deuterium content to be matched in all the considered discharges. We also compare the predicted and observed ratio between the neutron rate of the two pulses where neutron measurements are available. After adding the contribution from the beam - beam reactions, we obtain $ \frac{neutrons_{91227}}{neutrons_{91232}} \sim 3.4 $, similar to the experimentally measured value of $ \sim 3.5 $

\begin{table*}
\caption{Density peaking and isotope composition for the three pulses. Density peaking is defined as $ \frac{n_{\rho = 0} - n_{\rho = 0.8}}{n_{\rho = 0.8}} $. The modelled values are showed at the boundary conditions ($ \rho = 0.8 $), the limit of validity of the simulations ($ \rho = 0.3 $) and the axis ($ \rho = 0.0 $). The experimental value is intended on axis, where $ n_{D}(r)/n_{e}(r) = (1+(1-r/a)\cdot\alpha)\cdot(n_{D}/n_{e})_{edge} $ was assumed and $ \alpha $ was varied to match the experimental neutron rate \cite{Maslov2018}}.
    \begin{indented}
    	\centering
    	\item[]\begin{tabular}{cccccc}
    	    \br
    		& \multicolumn{4}{c|}{Simulation} & Experiment \\ 
    		\mr
    		Pulse $ \# $ & $ \frac{n_{e(\rho = 0)} - n_{e(\rho = 0.8)}}{n_{e(\rho = 0.8)}} $ & $ \frac{n_{D}}{n_{H} + n_{D}} $ at $ \rho = 0.8 $ & $ \frac{n_{D}}{n_{H} + n_{D}} $ at $ \rho = 0.3 $ & $ \frac{n_{D}}{n_{H} + n_{D}} $ at $ \rho = 0 $ & $ \frac{n_{D}}{n_{H} + n_{D}} $ exp\\
    		\mr
    		91754 & 0.98 & 0.395 & 0.39 & 0.365 & - \\

    		91232 & 0.85 & 0.145 & 0.170 & 0.205 & 0.18 \\

    		91227 & 0.73 & 0.650 & 0.645 & 0.660 & 0.676 \\
    		\br
    	\end{tabular}
	\label{tab:experiment_vs_model_density}
	\end{indented}
\end{table*}

\section{Sensitivities} \label{Sensitivities}

Sensitivity scans were performed to show that the phenomena under consideration are not dependent on any of the modelling assumptions. This also allows us to investigate potential reasons for discrepancies between the model results and the experimental data. The impacts of boundary conditions, codes used, physics assumptions, and modelling choices are presented here.

As anticipated in \sref{Boundary}, the largest sensitivity regarding the boundary conditions is on the $ T_{i}/T_{e} $ ratio \cite{ho2019}. Larger $ T_{i}/T_{e} $ destabilizes ETG and stabilizes ITG \cite{Linder2019}, decreasing the particle outward diffusion. In this case where the NBI particle source is important, this translates in a more peaked density for higher $ T_{i}/T_{e} $. The electron temperature profiles are slightly changed through the collisional coupling, while the ion temperature profiles do not change significantly, due both to stiffness and the destabilising effect of $n_e$ peaking, which compensates the ITG stabilising effect of increased $T_i/T_e$. 

\begin{figure*}
	\centering
	\includegraphics[width=1.0\linewidth]{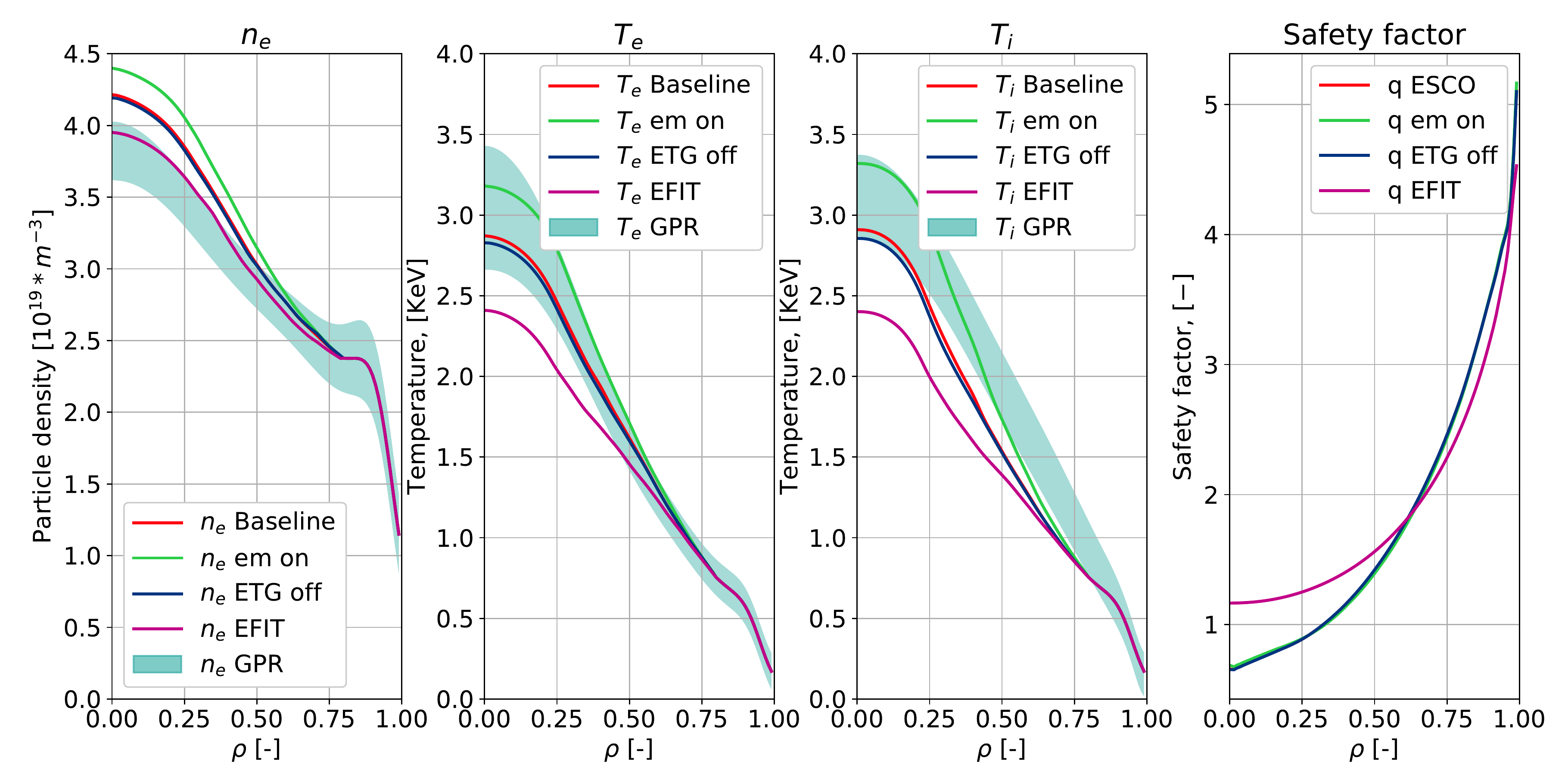}
	\caption{Effects of the boundary conditions on the profiles on pulse $ \#91227 $ (D beam, mixed puff). The green shaded area is the confidence interval given by the GPR, the solid lines the results from the modelling, with the red one being the baseline. $ T_{i} = T_{e} $ is imposed at $ \rho = 0.8 $}
	\label{fig:91227tite}
\end{figure*}

When $ T_{i} $ and $ T_{e} $ are changed at the same time, on the other hand, there is very little change on the temperature and density gradients. Modifying the density boundary condition also has little effect, rising the average density but without really impacting the gradients. The effect of the boundary conditions on the profiles is shown in \fref{fig:91227tite}. Regarding the inner ($ \rho <0.2 $) region where the modelling is not first-principle based, the profiles would be overestimated without adding extra transport. However, the difference is already small at $ \rho = 0.2 $ and disappears around $ \rho = 0.3 $. This has an effect on the value of $ \frac{n_{D}}{n_{H} + n_{D}} $ at the axis, as visible in the last panel in figure \ref{fig:Density_profiles}, where it changes a few percents around $ \rho = 0.2 $.

Changes in the total NBI power or in the ion - electron heating ratio, when not dramatic, proved not to be important in determining the final profiles. A sensitivity using the source profiles obtained from a static PENCIL calculation, with a $ \sim 7\% $ difference in the total particle source, showed very small differences with the Basecase. It was thus decided not to attempt more precise but time consuming source calculations using ASCOT, which is available in JINTRAC.  Regarding the edge neutrals source calculated with FRANTIC, even when raised to unrealistic values it could not change the $ \frac{n_{D}}{n_{H} + n_{D}} $ in the core by more than $ 1\% $. A doubling of the ionization energy and of the energy of the neutrals released from the wall also had negligible impact on the profiles. It has to be stressed, however, that the ratio is kept fixed at $ \rho = 0.8 $, where FRANTIC impact is already weak. An imprecision in the source calculations therefore does not impact our primary results.

\begin{figure*}
	\centering
	\includegraphics[width=1.0\linewidth]{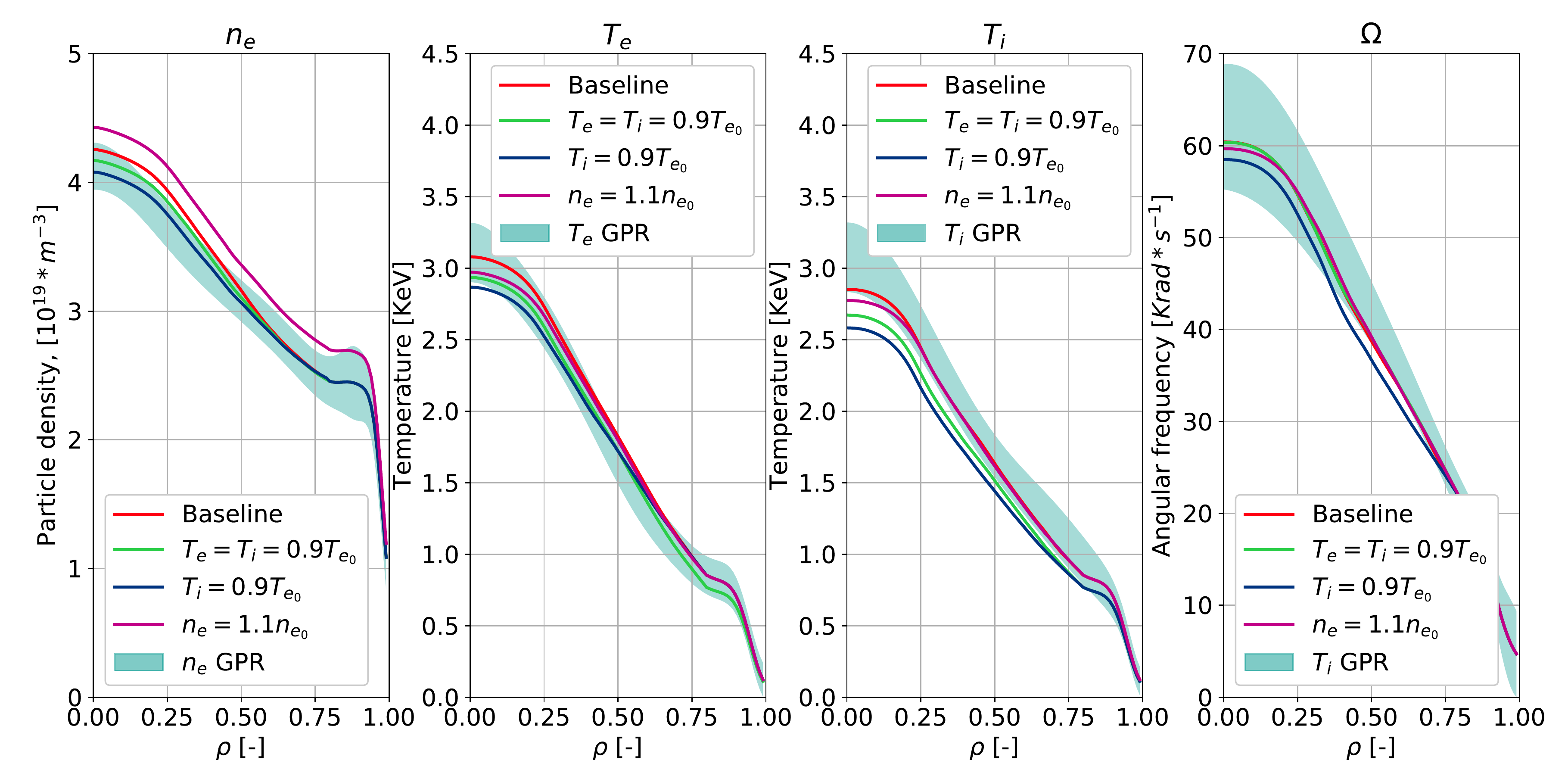}
	\caption{Physics sensitivities on pulse $ \#91232 $ (D beam, H puff). The green shaded area is the confidence interval given by the GPR, the solid lines the results from the modelling, with the red one being the baseline. The EFIT Q profile was taken from the inter-shot EFIT and the simulation was run with interpretive current}
	\label{fig:91232sensitivities}
\end{figure*}

\begin{table*}
\caption{The most important sensitivities of the profiles against the modelling assumptions. The values refers to pulse $ \# $91227 (D beam, mixed puff). Very similar results were obtained for the other two pulses and are not reported for brevity.}
    \begin{indented}
	    \centering
	    \item[]\begin{tabular}{cccccc}
		    \br
		    Simulation & $ \langle n_{e} \rangle [ m^{-3}10^{19} ] $ & $ T_{e, \rho=0} $ [KeV] & $ T_{i, \rho=0} $ [KeV] & $ \frac{n_{\rho = 0} - n_{\rho = 0.8}}{n_{\rho = 0.8}} $ & $ \frac{n_{D}}{n_{H} + n_{D}} $ at $ \rho = 0.3 $  \\
    		\mr
    		Baseline & 2.8 & 3.35 & 3.50 & 0.62 & 0.63 \\ 
    		
    		$ T_{i, \rho = 0.8} + 10\% $ & 2.9 & 3.30 & 3.40 & 0.80 & 0.64 \\
    		
    		$ T_{i, \rho = 0.8} - 10\% $ & 2.7 & 3.40 & 3.50 & 0.5 & 0.61 \\ 
    		
    		$ n_{e, \rho = 0.8} + 10\% $ & 3.0 & 3.25 & 3.35 & 0.58 & 0.63 \\ 
    		
    		em off & 2.6 & 3.15 & 3.0 & 0.44 & 0.61 \\ 
    		
    		ETG off & 2.7 & 3.50 & 3.50 & 0.48 & 0.61 \\ 
    		
    		Rotation off & 2.6 & 3.30 & 3.40 & 0.46 & 0.62 \\ 
    		
    		No Impurities & 2.8 & 3.10 & 3.20 & 0.64 & 0.63 \\ 
    		
    		EFIT $q$-profile & 2.7 & 2.85 & 2.90 & 0.52 & 0.625 \\ 
    		\br
    	\end{tabular}
    	\label{tab:sensitivities}
    \end{indented}
\end{table*}

$ Z_{eff} $ was very low in all the cases considered, below 1.2, but it is still important to consider impurities. Higher $ Z_{eff} $ is stabilizing for ITG/TEM modes due to the increase in collisionality and dilution, and also increases the ETG critical gradient threshold \cite{Jenko2001}, so even for these low impurity concentration there is an impact on the electron temperature. Impurities are to be considered in the comparison with the experiment, but it is important to notice that nor dilution nor the effect on $ T_{e} $ were found to modify the $ \frac{n_{D}}{n_{H} + n_{D}} $ ratio.
The radiation can in principle modify the density peaking, but only when the levels of radiation are comparable with the electron heating, which was not the case in the experiments that were considered. Doubling the radiation had very little effect on the profiles, thus justifying the choice not to include W in the modelling. The rotation was also found not to have a large impact on the profiles. The largest sensitivity is on $ \#91232 $ (D beam, H puff), changing the density on axis by $ \sim 5\% $ and the ion temperature by $ \sim 8\% $.

Electromagnetic (EM) ITG stabilization effects \cite{Citrin2013, Pueschel2010} are not self-consistently included in QuaLiKiz. Since the expected level of EM-stabilization is strongly correlated with the fast ion content in discharges with strong NBI heating \cite{Citrin2015a}, an ad-hoc correction can be applied to mimic this effect. The normalized logarithmic ion temperature gradient passed to QuaLiKiz, $ R/L_{T_{i}} $, is multiplied by the ratio of the local thermal energy density over the local total energy density, $ W_{th}/W_{tot}$. This simple ad-hoc model has been shown to consistently improve ITG predictions in high-$\beta$, high performance hybrid scenarios \cite{Casson2020}. This ad hoc model, due to the large fraction of fast-ion pressure ($ \sim 25\% $ in the D-beam heated pulses), also has a strong impact on the profiles for our cases. The ion temperature is increased, and given the relatively high collisionality the electron temperature increases as well. The $T_{i}/T_{e} $ ratio also changes, and this modifies the density peaking. There is in this case an effect on the $ \frac{n_{D}}{n_{H} + n_{D}} $ ratio, since it weakly correlates with the density peaking. However, this weak effect on the ratio is not nearly large enough to contradict the main point. We stress that the applicability of the ad-hoc model for our low-$\beta$ cases is not necessarily valid, hence the decision not to include it in the Basecase but to only study the sensitivity. Further investigation with high-fidelity gyrokinetic modelling is possible, but out of the scope of this work.

Including the ETG scales is not seen to have a large impact on the profiles, suggesting a low ETG activity in these pulses. Interestingly, the ETG impact is found to be stronger when used in combination with the ad-hoc em stabilizer. This is in line with the observation of ETG being important in hybrid scenarios.

The current profile also has a strong effect on the kinetic profiles. The sensitivity from the point of view of integrated modelling is shown in figure \ref{fig:91232sensitivities}. The q-profile that results from unconstrained EFIT, which was used exploiting its inaccuracy to check the sensitivity on the q-profile, is overestimated in the core. Assuming similar boundary conditions, as is the case, this also means underestimated magnetic shear. In the simulation where the current is predicted, on the contrary, the magnetic shear monotonically increases as the simulation evolves. Since at relaxation the value for the internal inductance is greater than the experimental one, we can assume a slight overestimation of the current peaking, and therefore of the magnetic shear. The two cases can be taken as two extreme cases.

The pulse with the largest variation between edge and core D concentration is, unsurprisingly, 91232. In this case there is no Deuterium puffing and its concentration is at the lowest. The relative intensity of the source is the largest when compared to the Deuterium density, so a larger impact is expected.

The kinetic profiles can have rather strong dependencies on the assumptions, around $ 20\% $ in the most different cases for the central values.  The $ n_{D}/(n_{D}+n_{H}) $ ratio, instead, never changes more than a couple of percentage points. The density peaking itself can change considerably, as is shown in \tref{tab:sensitivities}. Even when this happens, once the ratio is fixed at the edge it is very difficult to change it, at least in the model. Both Deuterium and Hydrogen are found to be peaked in all cases, since the peaking depends on the fact that ITG is unstable, and this does not change between the various simulations. The phenomena is robust against the assumptions.

\begin{figure}
	\centering
	\includegraphics[width=1.0\linewidth]{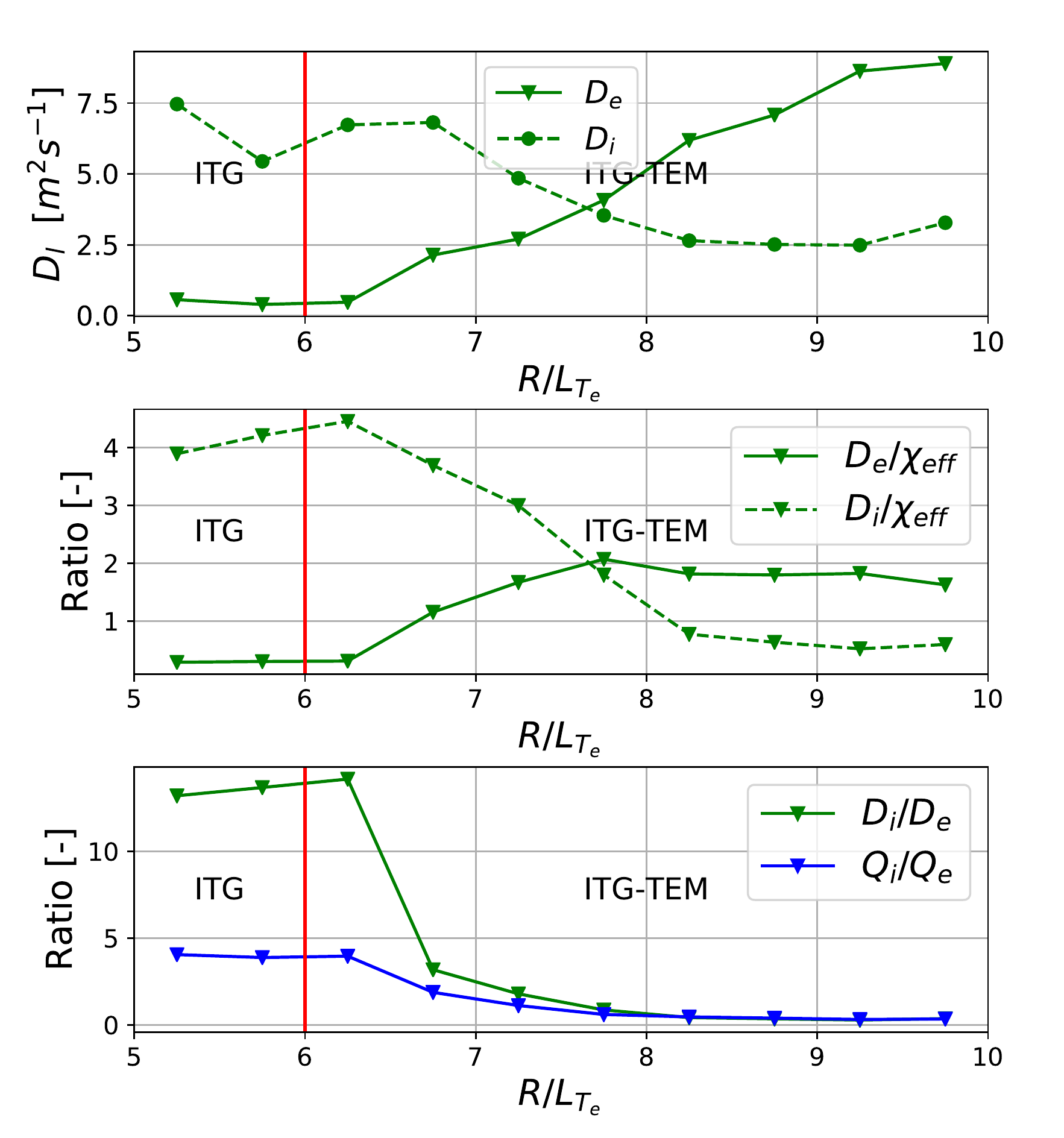}
	\caption{Top plot: particle transport coefficients for a $ R/L_{T_{e}} $ scan, based on the last timestep of the Basecase, shot $ \#91232 $ (D beam, H puff), $ \rho = 0.5$, with decreased collisionality and $R/L_{T_i}$ to accentuate an ITG-TEM transition. Points and triangles represent ion and electron particle diffusivity. Central plot: ratio between the electron (circles) and ion (triangles) diffusion coefficients and $ \chi_{eff} $. Bottom plot: ratio between ion and electron particle transport coefficients (green) and heat fluxes (blue)}
	\label{fig:ITG_to_TEM_2}
\end{figure}

\begin{figure}
	\centering
	\includegraphics[width=0.9\linewidth]{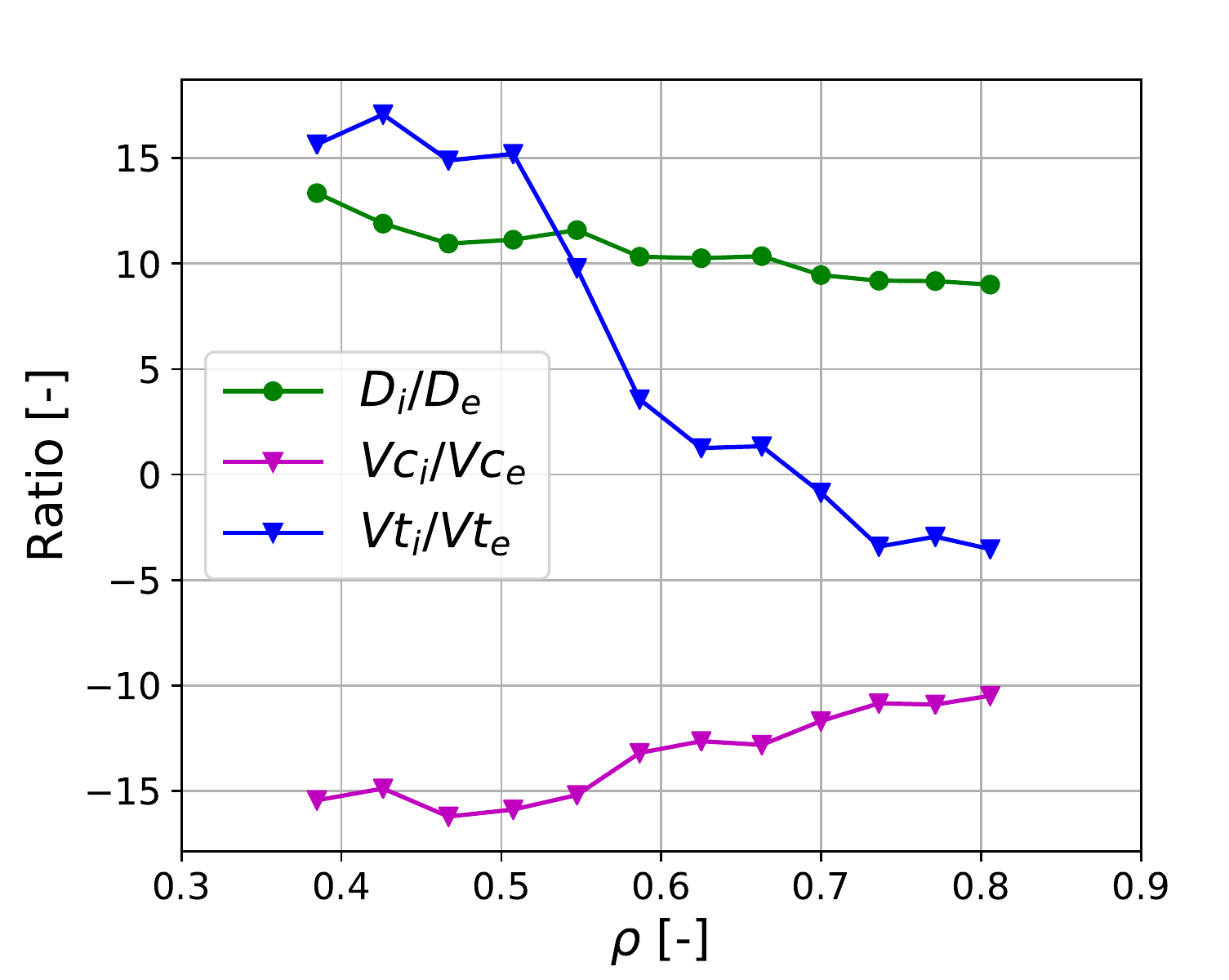}
	\caption{Ratio between the ion and the electron particle transport coefficient as a function of the radius. $ D_{s} $ is the diffusivity of the specie s, $ Vt_{s} $ the thermodiffusion and $ Vc_{s} $ the pure convective term. The values are taken from the QuaLiKiz calculations at the last timestep of the Basecase, shot $ \#91232 $ (D beam, H puff). No turbulence was found unstable inside $ \rho = 0.3 $}
	\label{fig:DiDe}
\end{figure}

Our key conclusions were found experimentally and via modelling in an ITG dominated regime. It is paramount to investigate the robustness of the effect in a more reactor relevant regime, where $Q_i \sim Q_e$, and a mixture of ITG and TEM instabilities are expected \cite{fable2019}. This was investigated using standalone QuaLiKiz, and shown in fig. \ref{fig:ITG_to_TEM_2}. An increasing $ R/L_{T_{e}} $ scan was carried out, where TEM is progressively destabilized. 
The input parameters are as in figure \ref{fig:DiDe}, but where $R/L_{T_i}$ (by 15\%) and collisionality (by factor 10) were reduced to approximate more reactor-relevant conditions. A transition from pure ITG to a regime with coexisting ITG-TEM was found at reasonable $R/L_{T_{e}}$. Only ion-scale modes were considered. Critically, in the $ Q_{i}/Q_{e}\approx1$ reactor-relevant regime in the vicinity of $R/L_{T_e}\approx8$, $D_e$ rises significantly, and $ D_{i} $ still remains large. Similar results are observed for the absolute values of the convective terms, but not shown for brevity. These observations suggest that the conclusions of this work remain valid for reactors. Isotope profiles tied to electron profiles regardless of core isotope source, fast isotope-mixing, and electron density peaking dominated by transport (due to large $D_e$~\cite{Tala2019}), can all co-exist in a reactor relevant regime.

\section{Conclusions} \label{Conclusions}

This work directly follows from experimental \cite{Maslov2018} and theoretical \cite{Bourdelle2018a} observation of large ion particle transport coefficients for ITG dominated plasmas. In particular, the focus was on a set of multi-isotope (D+H) experiments at JET that measured the isotope profiles while changing the relative magnitude of the edge and core D+H particle sources. Insensitivity of the shape of the isotope profiles to their respective sources was observed. The quasilinear turbulent transport model QuaLiKiz was shown to capture this effect in flux-driven modelling within the JINTRAC integrated modelling suite. 
Three discharges with different isotope compositions and core and edge isotope sources were modelled with predictive $ j $, $ n_{e} $, $ n_{H} $, $ n_{D} $, $ n_{Be} $, $ T_{e} $, $ T_{i} $, and $ \Omega $. Good correspondence with the experimental measurements was obtained. This increases the set of validated discharges for multi-channel integrated modelling with QuaLiKiz. Furthermore, in all analyzed pulses, the same key prediction was obtained: the isotope profile was found to be determined by the edge composition and the electron profile and to be weakly sensitive to the core particle source.

Extensive sensitivity analysis was performed, varying the boundary conditions, physical assumptions and codes used. Differences in temperature profiles and in the density peaking were found between the various cases. But the primary observation regarding the core isotope composition was robust against these changes, since it mostly depends on ITG being unstable.

These results provide confidence in predictive QuaLiKiz modelling in multi-isotope regimes, and extend the predictions to different machines or regimes, including the JET DT campaign. For example, this validation increases the confidence on the  multiple-isotope behaviour for the full-power DT extrapolation performed in \cite{Casson2020, Garcia2019}.

In a mixed ITG-TEM regime, the expectation - as also seen in QuaLiKiz standalone studies - is that both ion and electron transport coefficients are large, around $ 2 \cdot \chi_{eff}$, with a relatively weak impact of source on all channels. This is the predicted turbulence regime of reactors, where the combination of dominant electron heating and ion-electron heat exchange lead to $Q_i \sim Q_e$ \cite{fable2019} and significant density peaking with no source.

The results have positive ramifications for multiple-isotope fuelling, since for ITG or mixed ITG-TEM plasmas, controlling the isotope composition at the edge will be enough to control the composition in the core. Furthermore, the large ion transport coefficients imply fast relaxation of the isotope composition following transients, which is key for reactor burn-control applications and core He ash removal. This regime is ITER and reactor relevant, where ITG is predicted to be unstable due to ion-electron heat exchange, in spite of primarily electron heating \cite{FKoechlIAEA2018}.

\section{Acknowledgments} \label{Acknowledgments}

This work has been carried out within the framework of the EUROfusion Consortium and has received funding from the Euratom research and training programme 2014-2018 and 2019-2020 under grant agreement No 633053. The views and opinions expressed herein do not necessarily reflect those of the European Commission.

\printbibliography

\end{document}